# Improved QPP Interleavers for LTE Standard

Lucian Trifina[1], Daniela Tărniceriu[1], Valeriu Munteanu[1]

[1] "Gheorghe Asachi" Technical University of Iași, Electronics, Telecommunications and Information Technology Faculty, Department of Telecommunications, Bd. Carol I, no. 11, Iași, 700506, Romania,
luciant@etti.tuiasi.ro

*Abstract*— **This paper proposes and proves a theorem which stipulates sufficient conditions the coefficients of two quadratic permutation polynomials (QPP) must satisfy, so that the permutations generated by them are identical. The result is used to reduce the search time of QPP interleavers with lengths given by Long Term Evolution (LTE) standard up to 512, by improving the distance spectrum over the set of polynomials with the largest spreading factor. Polynomials that lead to better performance compared to LTE standard are found for several lengths. Simulations show that 0.5 dB coding gains can be obtained compared to LTE standard.**

**Keywords: QPP interleaver, spreading, distance spectrum, turbo codes.**

## I. INTRODUCTION

The selection of turbo coding was considered during the study phase of Long Term Evolution (LTE) standard to meet the stringent requirements (packet data support with data rates up to 100 Mbps on the downlink and 50 Mbps on the uplink, a low latency of 10 ms layer-2 round trip delay, flexible bandwidths up to 20 MHz, improved system capacity and coverage, and efficient VoIP support). Following deliberations within the working group, the quadratic permutation polynomials (QPP) were selected as interleavers for turbo codes, emerging as the most promising solutions to the LTE requirements. The QPP interleavers for the LTE standard [1] involve 188 different lengths. In this paper we intend to find QPP interleavers leading to improved distance spectrum.

The polynomial interleavers offer the following benefits [2]: special performance, complete algebraic structure, and efficient implementation (high speed and low memory requirements).

A QPP interleaver of length $L$ is defined in [2] as:
$$\pi(x) = (q_0 + q_1 x + q_2 x^2) \bmod L, x = \overline{0, L-1} \quad (1)$$
where $q_1$ and $q_2$ are chosen so that the quadratic polynomial in (1) is a permutation polynomial (i.e. the set $\{\pi(0), \pi(1), \ldots, \pi(L-1)\}$ is a permutation of the set $\{0, 1, \ldots, L-1\}$) and $q_0$ determines a shift of the permutation elements.

In the following we only consider quadratic polynomials with free term $q_0 = 0$, as for the QPP interleavers in the LTE standard. If $\mathbb{Z}_L = \{0, 1, \ldots, L-1\}$, then the permutation function is $\pi : \mathbb{Z}_L \to \mathbb{Z}_L$.

The spread factor $D$ is defined as [2]
$$D = \min_{\substack{i \neq j \\ i,j \in \mathbb{Z}_L}} \{\delta_L(p_i, p_j)\}. \quad (2)$$

$\delta_L(p_i, p_j)$ is the Lee metric between points $p_i = (i, \pi(i))$ and $p_j = (j, \pi(j))$:
$$\delta_L(p_i, p_j) = |i-j|_L + |\pi(i) - \pi(j)|_L, \quad (3)$$
where
$$|i-j|_L = \min\{(i-j)(\bmod L), (j-i)(\bmod L)\}. \quad (4)$$

The quadratic polynomials which lead to the largest spreading factor $D$ for some interleaver lengths are given in [2]. An algorithm for faster computation of $D$ is also presented. It is based on the representatives of orbits in the representation of interleaver-code.

Section II proposes a sufficient condition for two QPP interleavers to be identical. In Section III we present a method that leads to QPP interleavers with improved distances spectra compared to those given by the LTE standard. A table with these better QPP polynomials found for lengths up to 512 is given. Section IV presents the frame error rates (FER) resulted from simulations for LTE standard interleavers and the proposed ones for two different lengths. Section V concludes the paper.

## II. SUFFICIENT CONDITION FOR IDENTICAL QPP INTERLEAVERS

In this section we present a theorem which states sufficient conditions to be satisfied by the coefficients of two quadratic polynomials, so that the resulting interleavers are identical. For the LTE standard, the interleaver's length is always an even number.

*Theorem*

Consider two QPP interleavers described by the following polynomials (the free term is considered zero):
$$\pi_1(x) = (p_1 x + p_2 x^2) \bmod L, x = 0, 1, \ldots, L-1 \quad (5)$$
$$\pi_2(x) = (q_1 x + q_2 x^2) \bmod L, x = 0, 1, \ldots, L-1 \quad (6)$$

If $L$ is even, then $p_1 > q_1$, and the following relation is fulfilled:

$$p_1 - q_1 = \pm(p_2 - q_2) = L/2, \quad (7)$$

the two quadratic polynomials lead to identical permutations.

*Proof*:

For the two QPP to lead to identical permutations, it is required that

$$\pi_1(x) = \pi_2(x), \forall x = 0, 1, ..., L-1 \quad (8)$$

We denote

$$p_1 x + p_2 x^2 = k_1 \cdot L + \pi_1(x) \quad (9)$$

$$q_1 x + q_2 x^2 = k_2 \cdot L + \pi_2(x) \quad (10)$$

where $k_1, k_2 \in \mathbb{N}$. Under the conditions above we have to show that there are $k_1, k_2 \in \mathbb{N}$, $\forall x=0, 1, ..., L-1$, which verify relationship (8). Subtracting (10) from (9) and considering (8), we have:

$$(p_2 - q_2) x^2 + (p_1 - q_1) x = (k_1 - k_2) \cdot L \quad (11)$$

The solution of this quadratic equation is:

$$x = \frac{-(p_1 - q_1) + \sqrt{(p_1 - q_1)^2 + 4(p_2 - q_2)(k_1 - k_2)L}}{2(p_2 - q_2)} \quad (12)$$

Using (7) from the theorem statement, we have

$$x = \frac{-\frac{L}{2} + \sqrt{\left(\frac{L}{2}\right)^2 \pm 4\left(\frac{L}{2}\right)(k_1 - k_2)L}}{\pm L} = \frac{-1 + \sqrt{1 \pm 8(k_1 - k_2)}}{\pm 2} \quad (13)$$

where

$$k_1 - k_2 = \frac{1 - (-2x+1)^2}{8} = -\frac{x(x-1)}{2} \quad (14)$$

or

$$k_1 - k_2 = \frac{(2x+1)^2 - 1}{8} = \frac{x(x+1)}{2}. \quad (15)$$

Relationships (14) and (15) could also be obtained as follows. From (7) we have

$$p_1 = q_1 + L/2 \quad (16)$$

$$p_2 = q_2 \pm L/2 \quad (17)$$

Then, from (9) we get

$$k_1 \cdot L = p_1 x + p_2 x^2 - \pi_1(x) \quad (18)$$

or, taking into account (16), (17) and (8),

$$k_1 \cdot L = q_1 x + q_2 x^2 - \pi_2(x) + (L/2) \cdot (x \pm x^2) \quad (19)$$

or, from (10),

$$k_1 \cdot L = (L/2) \cdot (x \pm x^2) + k_2 \cdot L \quad (20)$$

From here, (14) and (15) result immediately.

Because $\forall x = 0, 1, ..., L-1$, $x(x-1)$ and $x(x+1)$ are divisible by 2, i.e. $k_1 - k_2 \in \mathbb{Z}$ then, there are $k_1, k_2 \in \mathbb{N}$ which verify relation (8).

The theorem shows that two QPPs generate identical permutation functions, if the coefficients are at the same distance ($L/2$). Therefore, we can only consider coefficients $q_1 = 0, 1, ..., (L/2) - 1$ in polynomial searching, because the interleaver $\pi(x) = (q_1 x + q_2 x^2) \bmod L$ is the same as the interleaver $\pi(x) = ((q_1 + L/2) x + (q_2 + L/2) x^2) \bmod L$, if $q_2 < L/2$, or as the interleaver described by the permutation $\pi(x) = ((q_1 + L/2) x + (q_2 - L/2) x^2) \bmod L$, if $q_2 \geq L/2$. As the number of searched QPPs is halved, so is the search time, therefore speeding up the search process.

### III. QPP INTERLEAVERS WITH IMPROVED DISTANCE SPECTRUM FOR LTE STANDARD

The search method of QPP interleavers consists firstly in selecting polynomials with maximum $D$. Among these, those with the best distance spectrum are chosen. The method is similar to method 2 in [3], where in the second step polynomials with the highest minimum distance and lowest multiplicities were chosen. Independent Rayleigh fading channel with known channel state information is considered. The used error measures are the truncated upper bounds (TUB) of bit error rates (BER) and of frame error rates (FER) [4]:

$$TUB(BER) = 0.5 \cdot \sum_{i=1}^{M} \frac{w_i}{L} \cdot \left(\frac{1}{1 + R_c \cdot SNR}\right)^{d_i}, \quad (21)$$

$$TUB(FER) = 0.5 \cdot \sum_{i=1}^{M} N_i \cdot \left(\frac{1}{1 + R_c \cdot SNR}\right)^{d_i}, \quad (22)$$

where $M$ is the number of terms in the distance spectrum taken into account, $d_i$ is the $i^{th}$ distance in the spectrum, $w_i$ is the total information weight corresponding to distance $d_i$, $N_i$ is the number of code words with distance $d_i$, $R_c$ is the coding rate and SNR is the signal to noise ratio.

These upper bounds of BER and FER are true for high SNR and maximum likelihood decoding. However, it was shown [8] that for high SNR, BER and FER values for suboptimum decoding of turbo codes converge to BER and FER values corresponding to maximum likelihood decoding. Thus, these bounds can be used to assess the performances of turbo codes with different interleavers in error-floor region.

The method selects the QPPs which maximize $D$. From the obtained set, those leading to minimum TUB(FER) are chosen, because FER is of more interest to wireless transmissions. These interleavers are denoted by LS-QPP-TUB(FER)min, where LS stands for largest spread. In both searches the result from Section II is used, which leads to halving the number of polynomials for which parameter $D$ and the distance spectrum are calculated. To calculate the distance spectrum, we used Garello's method [5], [6]. Because the number of terms in the spectrum is greater than 1, we cannot use the same value for the parameter wu_max as in [3]. This value is set to 10, to lead to the exact computed distance spectrum, as given in [6]. To reduce the computing time, we can reduce the number of terms of the spectrum, when the length increases.

The SNR value decreases when the length of the interleaver increases too much, in order not to result too small values for TUB(FER), but same magnitude order for all the lengths.

The used trellis termination is as in [7], transmitting the termination bits of the second trellis. Since the turbo code uses

a component code with memory 3, the coding rate is calculated by:

$$R_c = \frac{L}{3 \cdot L + 12} \quad (23)$$

Table I gives the QPP polynomials in the LTE standard and those found out by optimizing the distance spectrum for Rayleigh fading channel. For the specified SNR values and the considered number of distances, the values $10^7 \cdot$ TUB(BER) and $10^5 \cdot$ TUB(FER) are given. The value of the parameter $D$, minimum distances ($d_{min}$) and their multiplicities ($N_1$ and $w_1$) for each QPP interleaver are also given. The penultimate column gives the number of polynomials which lead to the highest value of $D$ and minimum TUB(FER) for that length. The table only presents polynomials with the lowest $q_1$ and then with the lowest $q_2$. In the last column the ratio between the TUB(FER) for the LTE interleaver and that found by the proposed method is given. The values in the table do not reflect the real value of the ratio between simulated FER values, but still show a significant performance difference. Moreover, higher performance differences are noticeable at SNR values higher than those in Table I.

We considered more than one term in distance spectrum, because only the minimum distance and its multiplicities have proved to be insufficient in some cases, leading to polynomials with weaker performances.

For example, for interleaver's length equal to 40, searching the distance spectrum with only one term leads to the polynomial $\pi(x) = 19x + 30x^2$, for which the minimum distance is $d_{min}$=14 and the multiplicities are $N_1$=2 and $w_1$=4. This polynomial, for 9 terms in distance spectrum, leads to following BER(FER) upper bonds: TUB(BER)·$10^7$=8.3564 and TUB(FER)·$10^5$=0.8106, respectively, values that are higher than those for the polynomial given in Table I.

For 5 length values (72, 168, 368, 440, 464), the obtained ratio was less than 1, meaning that LTE interleavers lead to a better distance spectrum than those found here. For these lengths more extensive searches were carried out, imposing the minimum parameter $D$ of polynomials among which the search is performed to be that of polynomials given in LTE standard. The found polynomials are given in Table II.

The remaining lengths result in a ratio of TUB(FER) values less than 2, i.e. close performance, or even 1, that is, interleavers identical as polynomial or distance spectrum (for lengths 56, 80, 88, 112, 176, 208, 280, 312, 344, 376, 440).

## IV. SIMULATION RESULTS

Simulations were performed for interleaver lengths equal to 40 and 448. The component code is that considered in the LTE standard, i.e., given by the generator matrix G = [1, 15/13]. The decoding algorithm is MAP (Maximum A Posteriori) with a stopping criterion based on LLR module (Logarithm Likelihood Ratio). The maximum number of iterations is 12, and LLR threshold is 10. The same number of blocks of bits for each SNR value was simulated for each length.

Obviously, the imposed number of blocks increases with the SNR value. The simulations were performed for a channel with independent Rayleigh fading and additive white Gaussian noise (AWGN). The used modulation is BPSK (Binary Phase Shift Keying).

Fig. 1 presents FER (solid line) and TUB(FER) (dashed point line) curves for the LTE interleaver and for the proposed one with largest spread and minimum TUB(FER), for length 40.

The ratio calculated in the last column of Table I is 2.48. From Fig. 1 we note that FER for the proposed interleaver is clearly lower than for the LTE-QPP interleaver for SNR greater than 5.5 dB. For example, at SNR = 8 dB, the ratio between the FER values is 2.27.

For FER=$10^{-5}$, the coding gain of the interleaver we found is greater with approximately 0.5 dB than that in LTE standard.

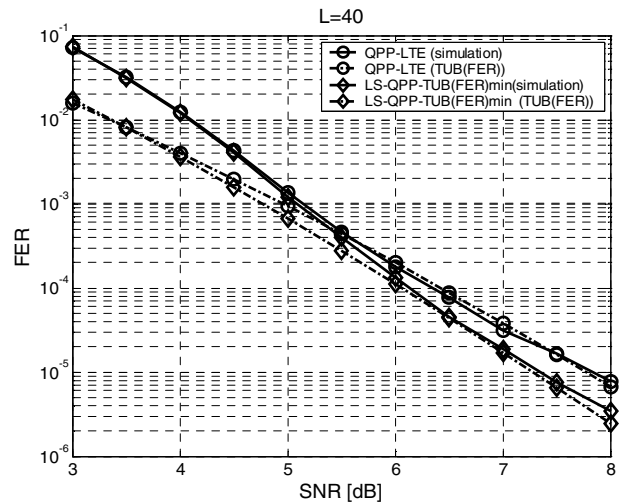

Figure 1. FER and TUB(FER) curves for LTE and the proposed LS-QPP-TUB(FER)min interleavers for length 40

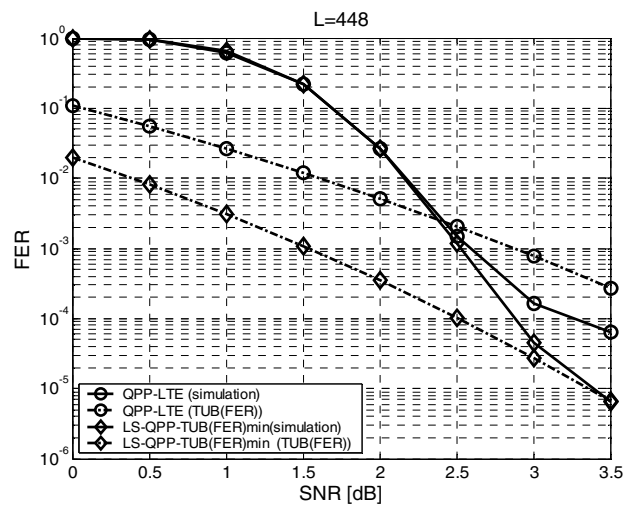

Figure 2. FER and TUB(FER) curves for QPP-LTE and the proposed LS-QPP-TUB(FER)min interleavers for length 448

TABLE I.  LTE-QPP AND LS-QPP-TUB(FER)min INTERLEAVERS

| L | SNR [dB] | num dist | LTE-QPP Interleavers | | | | | LS-QPP-TUB(FER)min Interleavers | | | | | No. pol. | FER_LTE/FERmin |
| | | | $\pi(x)$ | D | $d_{min}/N_1/w_1$ | TUB(BER) $*10^7$ | TUB(FER) $*10^5$ | $\pi(x)$ | D | $d_{min}/N_1/w_1$ | TUB(BER) $*10^7$ | TUB(FER) $*10^5$ | | |
|---|---|---|---|---|---|---|---|---|---|---|---|---|---|---|
| 40 | 7.5 | 9 | $3x+10x^2$ | 4 | 11/1/3 | 10.559 | 1.6211 | $13x+30x^2$ | 4 | 12/1/2 | 4.0451 | 0.6539 | 4 | 2.48 |
| 128 | 5.5 | 7 | $15x+32x^2$ | 16 | 16/12 | 1.2349 | 0.6560 | $17x+32x^2$ | 16 | 18/1/2 | 0.2189 | 0.1446 | 4 | 4.54 |
| 144 | 5 | 7 | $17x+108x^2$ | 16 | 20/2/4 | 0.4829 | 0.2873 | $19x+36x^2$ | 16 | 19/1/1 | 0.2131 | 0.1431 | 4 | 2.01 |
| 184 | 5 | 7 | $57x+46x^2$ | 12 | 16/1/2 | 1.0900 | 0.9958 | $25x+46x^2$ | 14 | 20/2/4 | 0.1083 | 0.0959 | 4 | 10.38 |
| 240 | 4.5 | 7 | $29x+60x^2$ | 16 | 24/2/4 | 0.5208 | 0.4262 | $89x+60x^2$ | 16 | 24/1/2 | 0.0897 | 0.0807 | 4 | 5.28 |
| 256 | 4.5 | 7 | $15x+32x^2$ | 16 | 16/1/2 | 1.3491 | 1.7748 | $31x+192x^2$ | 16 | 27/2/4 | 0.0131 | 0.0122 | 4 | 145.48 |
| 320 | 4 | 5 | $21x+120x^2$ | 20 | 20/1/2 | 0.3842 | 0.6802 | $21x+80x^2$ | 20 | 25/1/3 | 0.0209 | 0.0283 | 4 | 24.04 |
| 352 | 3.5 | 5 | $21x+44x^2$ | 22 | 20/1/2 | 1.1520 | 2.0785 | $153x+264x^2$ | 22 | 27/1/1 | 0.0291 | 0.0381 | 2 | 54.55 |
| 384 | 3 | 5 | $23x+48x^2$ | 24 | 22/1/2 | 1.0699 | 2.4172 | $25x+336x^2$ | 24 | 25/1/3 | 0.6408 | 1.0269 | 4 | 2.35 |
| 400 | 3 | 5 | $151x+40x^2$ | 16 | 19/1/1 | 1.1190 | 3.7777 | $47+100x^2$ | 20 | 24/1/2 | 0.8329 | 1.2787 | 4 | 2.95 |
| 408 | 3 | 5 | $155x+102x^2$ | 24 | 23/1/1 | 0.1678 | 0.5815 | $25x+306x^2$ | 24 | 27/2/4 | 0.1063 | 0.2099 | 4 | 2.77 |
| 416 | 3 | 5 | $25x+52x^2$ | 26 | 23/1/1 | 0.9586 | 1.8504 | $129+104x^2$ | 26 | 25/1/1 | 0.1133 | 0.3088 | 4 | 5.99 |
| 424 | 3 | 5 | $51x+106x^2$ | 24 | 24/1/2 | 0.2928 | 0.5601 | $157x+106x^2$ | 24 | 27/1/3 | 0.1220 | 0.2075 | 8 | 2.70 |
| 448 | 3 | 3 | $29x+168x^2$ | 28 | 22/105/210 | 34.639 | 77.621 | $139x+112x^2$ | 28 | 25/1/1 | 1.1863 | 2.7474 | 8 | 28.25 |
| 456 | 3 | 3 | $29x+114x^2$ | 24 | 23/1/1 | 0.1113 | 0.4657 | $55x+342x^2$ | 24 | 27/1/3 | 0.0402 | 0.0680 | 4 | 6.85 |
| 480 | 3 | 3 | $89x+180x^2$ | 30 | 26/2/4 | 0.1321 | 0.3745 | $209x+120x^2$ | 30 | 27/1/1 | 0.0220 | 0.0919 | 4 | 4.08 |
| 488 | 3 | 3 | $91x+122x^2$ | 24 | 27/2/4 | 0.0714 | 0.1743 | $181x+122x^2$ | 24 | 27/1/3 | 0.0440 | 0.0747 | 4 | 2.33 |
| 504 | 3 | 3 | $55x+84x^2$ | 28 | 29/1/1 | 1.5408 | 3.8857 | $197x+168x^2$ | 28 | 25/1/3 | 0.1288 | 0.2926 | 2 | 13.28 |

TABLE II.  LTE-QPP AND LS-QPP-TUB(FER)min INTERLEAVERS (MORE EXTENSIVE SEARCH)

| L | SNR [dB] | num dist | LTE-QPP Interleavers | | | | | LS-QPP-TUB(FER) min Interleavers | | | | | No. Pol. | FER_LTE/FERmin |
| | | | $\pi(x)$ | D | $d_{min}/N_1/w_1$ | TUB(BER) $*10^7$ | TUB(FER) $*10^5$ | $\pi(x)$ | D | $d_{min}/N_1/w_1$ | TUB(BER) $*10^7$ | TUB(FER) $*10^5$ | | |
|---|---|---|---|---|---|---|---|---|---|---|---|---|---|---|
| 368 | 3.5 | 5 | $81x+46x^2$ | 14 | 22/2/4 | 0.4270 | 0.7361 | $45x+92x^2$ | 16 | 28/1/4 | 0.0454 | 0.0337 | 8 | 21.84 |
| 464 | 3 | 3 | $247x+58x^2$ | 16 | 28/3/6 | 0.1053 | 0.1920 | $97x+116x^2$ | 20 | 29/1/1 | 0.0130 | 0.0407 | 4 | 4.72 |

We note that the curves corresponding to the simulation tend asymptotically to those corresponding to TUB(FER) bounds.

Fig. 2 shows the FER curves resulted from simulations and those corresponding to TUB(FER) bounds for length 448. We note the improved performance for the interleaver determined in Section III. At SNR = 3.5 dB, FER decreases by one order of magnitude. A coding gain of approximately 0.55 dB is obtained for FER = $6 \cdot 10^{-5}$.

V. CONCLUSIONS

The paper presents and proves sufficient conditions which have to be satisfied by the coefficients of two quadratic permutation polynomials so that the generated permutations are identical. The result is used to reduce by half the search time of QPP interleavers with lengths as in the LTE standard, by improving the distance spectrum over the set of polynomials with the largest spreading factor.

The method we proposed selects firstly the QPPs which maximize D and then, from the obtained set, those leading to minimum TUB(FER). This upper bound for FER was calculated for an independent Rayleigh fading channel.

Polynomials that lead to improved performances (TUB(FER) minimum) are found for several lengths.

The search time is significantly lower than that in the exhaustive search, because the search of TUB(FER) is performed only over the set of polynomials with maximum D.

The distance spectrum and TUB(FER) are calculated only for these polynomials.

The simulated FER curves outlined in Section 4 show that the interleavers found out by the proposed method can lead to superior performances compared to those in the LTE standard.

REFERENCES

[1] 3GPP TS 36.212 V8.3.0, 3rd Generation Partnership Project, Multiplexing and channel coding, 2008-05, (Release 8).
[2] Y. O. Takeshita, "Permutation Polynomial Interleavers: An Algebraic-Geometric Perspective," IEEE Trans. on Information Theory, vol. 32, no. 6, June 2007, pp. 2116-2132.
[3] D. Tarniceriu, L. Trifina and V. Munteanu, "About minimum distance for QPP interleavers," Annals of Telecommunications, vol. 64, no. 11-12, Nov.-Dec. 2009, pp. 745-751.
[4] J. Yuan, W. Feng and B. Vucetic, "Performance of Parallel and Serial Concatenated Codes on Fading Channels", IEEE Trans. on Communications, vol. 50, no. 10, Oct. 2002, pp. 1600-1608.
[5] R. Garello, P. Pierleoni, and S. Benedetto, "Computing the Free Distance of Turbo Codes and Serially Concatenated Codes with Interleavers: Algorithms and Applications," IEEE Journal on Selected Areas in Communications, vol. 19, no. 5, May 2001, pp. 800-812.
[6] http://www.tlc.polito.it/garello/turbodistance/turbodistance.html
[7] D. Divsalar and F. Pollara, "Turbo Codes for PCS Applications," Proceedings of ICC 1995, Seattle, WA., pp. 54-59, June 1995